\documentclass{desyproc}

\begin{document}
\title{Searching for neutrino oscillation parameters in long baseline experiments}

\author{{\slshape Sampsa Vihonen}\\[1ex]
University of Jyvaskyla, Department of Physics, P.O. Box 35, FI-40014 University of Jyvaskyla, Finland\\}

\contribID{8}

\confID{13130}
\desyproc{DESY-PROC-XXXX-XX}
\acronym{Magellan Workshop : Connecting Neutrino Physics and Astronomy}
\doi

\maketitle

\begin{abstract}
Developing neutrino astronomy requires a good understanding of the neutrino oscillations mechanism. The European strategy for neutrino oscillation physics sets a high priority on future long baseline neutrino experiments with the aim to measure the intrinsic parameters that govern the neutrino oscillations. In this work we take a look at the next generation of long baseline experiments and discuss their prospects in future research.
\end{abstract}

\section{Introduction}

The dream of using neutrino telescopes as messengers of the universe in the future is built on the idea that the neutrino oscillation mechanism is eventually resolved in the neutrino oscillation experiments. The prospects of these experiments include both determining the parameters of the standard three-neutrino paradigm and probing non-standard interactions (NSI) that may follow from potential Beyond Standard Model theories.

Neutrino oscillation experiments are typically categorized into solar, reactor, atmospheric and accelerator-based experiments. In this work we focus on long baseline neutrino experiments where an intense muon neutrino beam is directed to traverse long distances underground.

In the Standard Model of particle physics (SM) neutrino sector the relationship between the three neutrino mass states $(\nu_1,\nu_2,\nu_3)$ and the three flavor states $(\nu_e,\nu_\mu,\nu_\tau)$ is usually expressed as a $3\times 3$ matrix known as the PMNS matrix after Bruno Pontecorvo, Ziro Maki, Masami Nakagawa and Shoichi Sakata. The mixing of neutrino states is the source to the phenomenon that is known as neutrino oscillations. The PMNS matrix can be written in terms of the three mixing angles and one phase in the form \cite{Forero:2014bxa}:

\begin{eqnarray}\label{Eq:1}
U &=& \left( \begin{array}{ccc}
1 & 0 & 0 \\
0 & c_{23} & s_{23} \\
0 & -s_{23} & c_{23} \end{array} \right)
\left( \begin{array}{ccc}
c_{13} & 0 & s_{13}e^{-i\delta_\text{CP}} \\
0 & 1 & 0 \\
-s_{13}e^{i\delta_\text{CP}} & 0 & c_{13} \end{array} \right)
\left( \begin{array}{ccc}
c_{12} & s_{12} & 0 \\
-s_{12} & c_{12} & 0 \\
0 & 0 & 1 \end{array} \right)\nonumber\\
&&\times diag\left(e^{i\alpha_1/2},e^{i\alpha_2/2},1\right)
\end{eqnarray}

where $s_{ij} = \sin \theta_{ij}$ and $c_{ij} = \cos \theta_{ij}$ for $i,j=1,2,3$,
$\theta_{ij} \in [0, \pi/2]$ are the intrinsic mixing angles, $\delta_{CP} \in [0, 2\pi]$ is the Dirac CP violation phase and $\alpha_1,\alpha_2$ are two Majorana CP violation phases.

The transition probability between two neutrino flavor states $\nu_l \rightarrow \nu_{l'}$ is calculated as:

\begin{equation}\label{Eq:2}
P_{l l'} = \sum_{j,k} U_{lj}^{*} U_{l'j} U_{lk} U_{l'k}^{*} e^{-iL \Delta m_{jk}^2 / 2E},
\end{equation}

where $L$ is the baseline length of the experiment and $E$ is the energy of the neutrino produced. Here neutrino mass squared differences $\Delta m_{jk}^2 = m_{j}^2-m_{k}^2$ for $j,k=1,2,3$ are calculated from the neutrino masses $m_1, m_2$ and $m_3$.

The expression in Eq. (\ref{Eq:2}) can be further simplified into

\begin{equation}\label{Eq:3}
P_{l l'} = \delta_{l l'} - \sum_{j,k;j>k} \left[4\sin^2\Delta_{kj} \operatorname{Re} W_{l l'}^{jk} - 2\sin2\Delta_{kj} \operatorname{Im} W_{l l'}^{jk} \right],
\end{equation}

where $W_{l l'}^{j k} = U_{lj}^{*} U_{l'j} U_{lk} U_{l'k}^{*}$ and $\Delta_{kj} = \Delta m_{kj}^2 L / 4E$. It is seen from expression (\ref{Eq:3}) that the oscillation probabilities can be tuned to desirable values by choosing $L/E$ conveniently.

In this paper, we analyze the potential of the planned long baseline neutrino oscillation experiment LBNO at two tasks. On the one hand we study its performance for resolving the intrinsic $\theta_{23}$ octant degeneracy, which will be described in Section 4, and on the other hand we study its ability to probe the matter NSI effects, which will be described in Section 5. That is, we derive numerically the values of $\theta_{23}$ for which the octant degeneracy that can be resolved and also calculate the new upper bounds that LBNO could give to $|\varepsilon_{\alpha \beta}^m|$ where $\alpha, \beta = e, \mu, \tau$.

\section{Oscillation parameters: A general outlook}

The present values of the oscillation parameters are given in Table 1. The values of the reactor and solar mixing angles ($\theta_{13}$ and $\theta_{12}$ respectively) have been determined in previous neutrino experiments with good precision, but the value of the atmospheric mixing angle $\theta_{23}$ has yet to be determined accurately. The other unknown quantities are the sign of $\Delta m_{31}^2$ and the amount of CP violation $\delta_\text{CP}$ in the leptonic sector.

\begin{table}[h!]
\begin{center}
\begin{tabular}{|c|c|c|}\hline
\rule{0pt}{3ex}Parameter & Value \\ \hline
\rule{0pt}{3ex}$\theta_{12}$ & $31.8^\circ$... $37.8^\circ$ \\ \hline
\rule{0pt}{3ex}$\theta_{13}$ & $7.7^\circ$... $9.9^\circ$\\ \hline
\rule{0pt}{3ex}$\theta_{23}$ & $38.8^\circ$... $53.3^\circ$\\ \hline
\rule{0pt}{3ex}$\Delta m_{21}^2$ & $(7.11$ ...$\,8.18)\cdot 10^{-5}\text{eV}^2$\\ \hline
\rule{0pt}{3ex}$|\Delta m_{31}^2|$ & $(2.20$ ...$\,2.65)\cdot 10^{-3}\text{eV}^2$ \\ \hline
\rule{0pt}{3ex}$\delta_{CP}$ & $0$ ...$\,360^\circ$ \\ \hline
\end{tabular}
\caption{\label{para1}Standard neutrino oscillation parameters \cite{Forero:2014bxa}.
For the unknown $\delta_{CP}$ we have denoted the values considered in numerical calculations,
and the extra phases $\alpha_{1,2}$ only come into play in double-beta decay experiments.}
\end{center}
\end{table}

\section{Simulation of long baseline neutrino experiments}

One of the most promising methods to determine the values of neutrino oscillation parameters is to use long baseline experiments which are designed to study neutrino oscillations by sending neutrinos through long distances underground, where the matter presence affects the oscillation probabilities. In this work we simulate a long baseline neutrino oscillation experiment that was described the LBNO design study (see Ref. \cite{Das:2014fja} for more details).

In LBNO the idea is to send neutrino and antineutrino beams, produced at the CERN SPS accelerator, towards the Pyh\"{a}salmi mine, located in central Finland at the distance 2288 km from CERN, where the neutrino fluxes would be measured by using a two-phase Liquid Argon Time Projection Chamber. According to the design study, the liquid argon tank would have been paired with a magnetized muon detector. In this plan, the size of the detector was planned to have 20 kton fiducial mass. The 0.75 MW neutrino beam would have been created with the CERN SPS facility.

The simulation is done by using the GLoBES software \cite{Huber:2004ka,Huber:2007ji}. GLoBES calculates the oscillation probabilities and the corresponding neutrino rates for any given set of oscillation parameter values. The standard set of the software simulates the neutrino propagation from source to detector and computes the standard matter interactions, and with a suitable extension also takes into account all matter-induced NSI effects in the propagation. The software computes $\chi^2$ distributions to compare different sets of oscillation parameter values.

We will determine for both mass hierarchies, the 1$\sigma$, 2$\sigma$ and 3$\sigma$ sensitivity limit of the angle $\theta_{23}$ that LBNO can achieve with 5+5 --years neutrino and antineutrino run, allowing the CP phase to vary in the range $0$ to $360^\circ$. We will also determine the new $90\%$ confidence level upper bounds that LBNO could set to the absolute values of $\varepsilon_{\alpha \beta}^m$ parameters that describe the strength of non-standard interactions between neutrinos and the matter.

\section{Determination of the $\theta_{23}$ octant}

The main reason behind the large ambiguity in the $\theta_{23}$ value can be traced to degeneracy between $\theta_{23}$ and its opposite octant $90^\circ - \theta_{23}$. In this section we show where the octant degeneracy originates and for what $\theta_{23}$ values it can be resolved in an LBNO--like experiment.

In long baseline experiments one is interested mainly in the oscillation channels $\nu_\mu \rightarrow \nu_\mu$ (disappearance channel) and $\nu_\mu \rightarrow \nu_e$ (appearance channel). 
For small quantities $\Delta_{21}^2/\Delta_{31}^2$ and $\sin \theta_{13}$, the $\theta_{23}$ dependence of the oscillation probability $P_{\mu \mu}$ is of the form $\sin^2 2\theta_{23} \sin^2 \Delta$ where $\Delta = \Delta m_{31}^2 L / 4E$. Since

\begin{align*}\label{Eq:4}
P_{\mu \mu}(90^\circ - \theta_{23}) = P_{\mu \mu}(90^\circ),
\end{align*}

the measurements of the channel $\nu_\mu \rightarrow \nu_\mu$ are not suitable for resolving the degeneracy.

In a future long baseline neutrino experiment this problem with $\theta_{23}$ octants is resolved by studying the electron appearance from $\nu_\mu \rightarrow \nu_e$ oscillations. The leading term of $P_{\mu e}$ has a dependence on $\theta_{23}$ that is not octant degenerate, and hence the $\nu_\mu \rightarrow \nu_e$ oscillations bring in the primary component to the determination of $\theta_{23}$ octant.

Using the GLoBES software, we simulated the hypothetical experiment that was described in the LBNO design study (see Ref. \cite{Das:2014fja} for more details). In the simulations we assume the same oscillation parameter values that are given in Table 1.

\begin{figure}[ht]\centerline{\includegraphics[width=\textwidth]{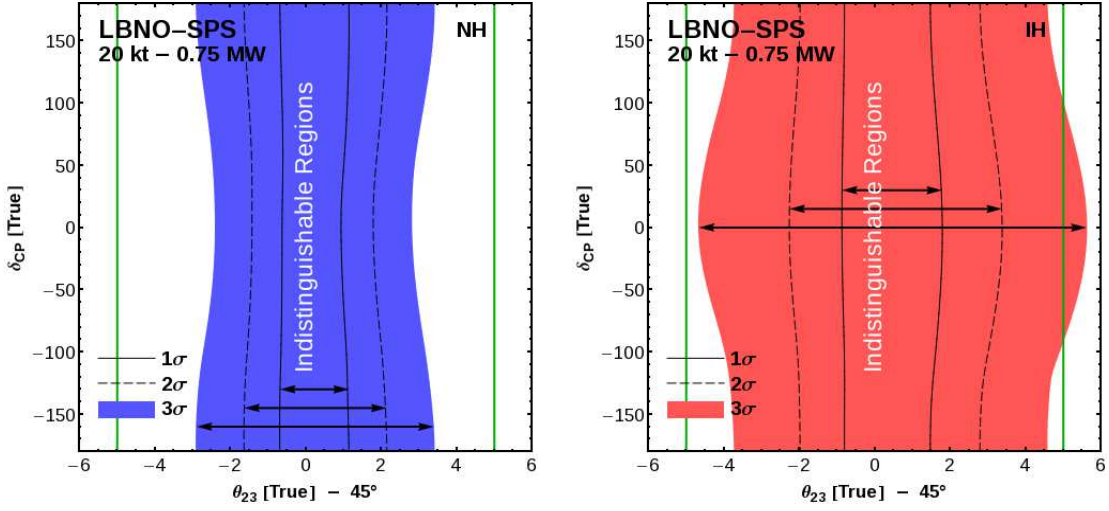}}
\caption{Determination ability of the $\theta_{23}$ octant in LBNO \cite{Das:2014fja} in normal hierarchy mode NH (left panel) and in inverted hierarchy mode IH (right panel). Straight vertical lines illustrate the $40^\circ$ and $50^\circ$ angles.}\label{Fig:1}
\end{figure}

The results are presented in Fig. \ref{Fig:1}. In both panels the white regions in the plots are the areas for which the values of $\theta_{23}$, $\delta_\text{CP}$ can be established with a confidence level greater than 3$\sigma$. Within these areas, one can eliminate with a confidence level larger than 3$\sigma$ the possibility for these parameters to lie in the other octant. The colored regions on the other hand show the values where no such distinction is possible with the indicated confidence level.

In Figure \ref{Fig:1} we have also marked the MINOS favored $\theta_{23}$ values 40$^\circ$ and 50$^\circ$ by green lines. It is seen that for this particular setup the right $\theta_{23}$ octant can be asserted in NH with at least 3$\sigma$ confidence level. As for IH this limit is reached for the lower octant, but for the higher octant it fails to be reached for $\delta_\text{CP}$ values between -100$^\circ$ and +100$^\circ$.

\section{Constraining non-standard interaction parameters}

In this section we analyze how an LBNO--like experiment would constrain the NSI effects that arise from extra couplings between neutrinos and matter.

The NSI effects can be parametrized in terms of effective charged current like (CC) and neutral current like (NC) Lagrangians \cite{Huitu:2016bmb}. In the low energy regime, these Lagrangians are given by

\begin{equation}\label{Eq:5}
\begin{split}
&\mathcal{L}^{CC}_{NSI} = -2\sqrt{2}G_F\varepsilon_{\alpha\beta}^{ff',C}(\overline{\nu}_{\alpha}\gamma^{\mu}P_L\ell_{\beta})(\overline{f}\gamma^{\mu}P_Cf'),\\
&\mathcal{L}^{NC}_{NSI} = -2\sqrt{2}G_F\varepsilon_{\alpha\beta}^{f,C}(\overline{\nu}_{\alpha}\gamma^{\mu}P_L\nu_{\beta})(\overline{f}\gamma^{\mu}P_Cf).
\end{split}
\end{equation}

In Equation (\ref{Eq:5}) the indices $f$ and $f'$ label charged leptons or quarks ($\ell_i$, $u_i$, $d_i$, $i = 1, 2, 3$), whereas $G_F = 1.166 \times 10^{-5} \text{GeV}^{-2}$ is the
Fermi coupling constant, $\alpha$, $\beta$ refer to neutrino flavor ($e, \mu, \tau$), and $C = L, R$ refers to the chirality structure of the charged lepton interaction. $P_L$ and $P_R$ are the
chiral projection operators. 

In this work we focus on the parameters $\varepsilon_{\alpha \beta}^{f f', C}$ and $\varepsilon_{\alpha \beta}^{f, C}$ which are dimensionless numbers. It is assumed here that the effective non-standard interactions have V--A Lorentz structure, and hence we allow both left-handed ($P_C = P_L$) and right-handed ($P_C = P_R$) couplings for the charged fermions. The charged current Lagrangian $\mathcal{L}^{CC}_{NSI}$ is relevant for the NSI effects in the source and detector, since both in the creation and detection processes involve charged fermions. The neutral current Lagrangian $\mathcal{L}^{NC}_{NSI}$ in turn is relevant for the NSI matter effects. The effective field theory described by Eq. (\ref{Eq:5}) are assumed to follow from some unspecified beyond-the-standard-model theory after integrating out heavy degrees of freedom.

Our interest is in the NSI effects that contribute to the propagation in matter. In this case, the NSI could contribute to the coherent forward scattering of neutrinos in the Earth’s crust. The effective Hamiltonian is given by:

\begin{equation}
H = \frac{1}{2E_{\nu}}\left[U
\left(
\begin{array}{ccc}
0 & 0 & 0 \\
0 & \Delta m_{21}^2 & 0\\
0 & 0 & \Delta m_{31}^2 
\end{array}
\right) U^{\dagger}
+V\right]
\label{Eq:6}
\end{equation}

where $E_\nu$ is the energy of the neutrino, $U$ is the mixing matrix (\ref{Eq:1}) and $V$ is the effective matter potential:

\begin{equation}
V = A\left(
\begin{array}{ccc}
1+\varepsilon_{ee}^m & \varepsilon_{e\mu}^m & \varepsilon_{e\tau}^m \\
\varepsilon_{e\mu}^{m*} & \varepsilon_{\mu\mu}^m & \varepsilon_{\mu\tau}^m \\
\varepsilon_{e\tau}^{m*} & \varepsilon_{\mu\tau}^{m*} &\varepsilon_{\tau\tau}^m
\end{array}\right).
\label{Eq:7}
\end{equation}

\begin{figure}[h!]
\centerline{\includegraphics[width=\textwidth]{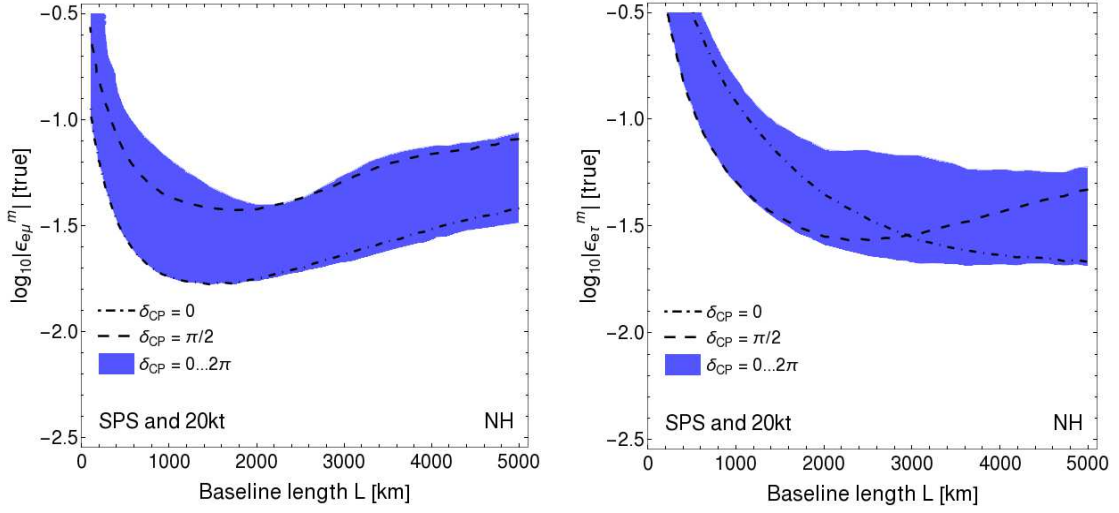}}
\caption{Constraining capability of LBNO for the absolute values of the non-standard interaction parameters $\varepsilon_{e \mu}^m$ and $\varepsilon_{e \tau}^m$ in normal hierarchy mode NH. See Ref. \cite{Huitu:2016bmb} for more details.}\label{Fig:2}
\end{figure}

Equations (\ref{Eq:6}) and (\ref{Eq:7}) include both the SM and NSI matter effects. We study numerically how the future neutrino oscillation experiments would constrain the absolute values of the NSI parameters $\varepsilon_{e \mu}$ and $\varepsilon_{e \tau}$. We will concentrate here on the future long baseline neutrino experiments, using the LBNO setup as our benchmark. For each $\delta_\text{CP}$ value, we have calculated the 90\% confidence level contour varying the baseline length parameter $L$ and the CP phase $\delta_\text{CP}$. In every case, a 90\% confidence level contour is found and the results merged in a contour band. The results are plotted in Fig. \ref{Fig:2}.

\section{Summary}

In this work we have presented the measurement potential of a potential future long baseline neutrino experiment in the determination of the $\theta_{23}$ octant and constrained the $| \varepsilon_{e \mu}^m|$ and $| \varepsilon_{e \tau}^m|$ parameter spaces for non-standard neutrino interactions that arise from the neutrino propagation in matter. In both cases we have demonstrated the interference both of these tasks suffer from the $\delta_\text{CP}$ parameter whose value is still unknown. We used the recent LBNO design study as our benchmark model and performed the numerical study with the GLoBES software.

\pagebreak
\begin{footnotesize}

\bibliographystyle{unsrt}
\bibliography{vihonen_sampsa}

\end{footnotesize}
\end{document}